\documentclass[twocolumn,14pt]{article}
\usepackage[dvips]{graphicx}
\begin{document}

\twocolumn [
\begin{center}
{\Large ADIABATIC AND ENTROPY PERTURBATIONS IN
INFLATIONARY  MODELS BASED ON NON-LINEAR SIGMA MODEL }
\end{center}
\bigskip

\textbf{N.A. Koshelev}$^1$ \\
\vskip 0.2cm {\it Ulyanovsk State University, 42 Leo Tolstoy St.,
Ulyanovsk 432700, Russia}

\begin{quote}
    The scalar perturbations in inflationary models, based on
    a two-component diagonal non-linear sigma model,  are considered.
    For inhomogeneities generated at an inflationary stage, the law
    of motion  of the comoving curvature ${\cal R}$ is obtained (without
    using the slow roll approximations). A formal expressions for
    the power spectrum and its spectral index are obtained, which are valid
    and after the slow-roll stage. As an example, an inflationary model with
    a massive scalar field and quadratic curvature corrections is studied
    numerically.
\end{quote}



\vskip 1cm  ] \footnotetext[1]{e-mail: koshna71@inbox.ru}

\section{Introduction.}

A standard hypothesis of modern cosmology is the existence of an
accelerated expansion stage in the history of the very early
Universe, when the second time derivative of the scale factor is
positive. This expansion was termed inflationary. Presence of a
long enough inflationary stage can resolve such well-known
difficulties of the Big Bang theory, as the homogeneity, horizon
and flatness problems \cite{Linde}. Besides, the inflationary
theory gives a root source of primordial inhomogeneities, giving
rise to the large-scale structure of the Universe. In the simple
models of chaotic inflation the existence only one scalar field
termed inflaton is supposed. In a wide range of initial values of
this field, the slow roll conditions is realized, and the scale
factor grows quasi exponential. The primordial density
inhomogeneities arise from quantum vacuum fluctuations of the
inflaton field. These quantum fluctuations, after horizon
crossing, can be considered as classical stochastic perturbations
\cite {PS}. At large scales, the conservation law  of the comoving
curvature perturbation $ \cal {R} $ allows to connect
inhomogeneities generated on inflationary stage with primordial
inhomogeneities in a radiation-dominated Universe. The slow-roll
models predict initial inhomogeneities with an approximately
scale-invariant (Harrison - Zel'dovich) spectrum in good agreement
with the astrophysical data \cite{Bridle}. However, more realistic
multi-field inflationary models also call the attention. In such
models, even for long-wavelength inhomogeneities, the conservation
law of the quantity $ \cal {R} $ may not to hold (for example,
strong amplification of long-wavelength inhomogeneities is
possible at the preheating stage). Many authors considered the
evolution $\cal {R}$ for a number of multi-field models both
analytically and numerically. For double-field inflationary
models, the equation of motion of the comoving curvature
perturbation was written out in the pioneering papers \cite{GW95},
\cite{GW96}.

In this work, we consider inflationary models based on a
non-linear sigma model with self-coupling potential. The general
two-component diagonal sigma-model action has the form:
\begin{eqnarray}
\label{e1}S = \int d^4 x\sqrt {|g|} \left\{  - \frac{R}{2
\kappa^2} + \frac{1}{2}T^2(\varphi, \chi)\varphi_{;\mu}
\varphi^{;\mu} \right. \nonumber\\
+\left. \frac{1}{2}P^2(\varphi,\chi) \chi_{;\mu}\chi^{;\mu} -
U(\varphi, \chi) \right\},
\end{eqnarray}

\noindent where the chiral metric components $T^2 (\varphi, \chi)
$, $P^2 (\varphi, \chi) $ are some functions of chiral fields
(sigma model components) $\varphi$ and $\chi$. The non-linear
sigma models usually arise after a conformal transformation in
generalized Einstein theories \cite{Maeda89}. In Refs.
\cite{Chervon}, \cite{CheKosh}, inflationary models with an action
of such view have received the name of chiral models. They are
often call simply multi-field models \cite{SaSt,Nibbelink},
however, we do not use this term to stress the non-triviality of
the scalar fields space.

In two-field inflationary models, to study adiabatic and entropy
perturbations, Gordon at al. \cite {GWBM}   constructed the
quantities $ \delta \sigma $ and $ \delta s $ . These variables
are very convenient for describing the time evolution of the
comoving curvature $ {\cal R} $, and for solving the perturbed
field and Einstein equations in slow roll. The perturbation
equations in the new variables are also suitable for numerical
simulations \cite {GWBM, TsuBassett}. This approach was
generalized to the case of a non-linear sigma model with chiral
metric components $h_{11}=1$, $h_{12}=0$, $h _ {22} = {\rm e} ^{2b
(\varphi_1)}$ \cite {MFB02}, and  to the case of the general
non-linear sigma model under slow roll conditions
\cite{Nibbelink}. The goal of the presented work is to study the
evolution of comoving curvature perturbations  and to obtain a
power spectrum of initial inhomogeneities in inflationary models
with the action (\ref{e1}).

\section{Background.}

We consider the spatially flat Friedmann - Robertson - Walker
(FRW) universe, described by the action (\ref{e1}) with the line
element
\begin{equation}
\label{e2}ds^2 =  dt^2 - a^2(t)\delta_{ij} dx^idx^j.
\end{equation}

The equations of motion for the homogeneous background chiral
fields $\varphi$, $\chi$ and scale factor $a(t)$ are
\begin{eqnarray}
\label{e3}H^2 = \frac {\kappa ^2} {3} \left ({ \frac{T^2}{2}
\dot{\varphi} ^2 + \frac{P^2}{2} \dot {\chi} ^2 + U} \right),  \\
\label{e4}\dot {H} = - \frac {\kappa ^2} {2} \left ({T^2 \dot
{\varphi} ^2 + P^2 \dot {\chi} ^2 } \right),
\end{eqnarray}
\begin{equation}
\label{e5}  \ddot{\varphi}  + 3H \dot{\varphi}
+2\frac{\dot{T}}{T}\dot{\varphi} - \frac{T_{,\varphi}}{T}
\dot{\varphi} ^2 - \frac{PP_{,\varphi}}{T^2} \dot{\chi} ^2
 + \frac{U_{,\varphi}}{T^2} = 0,
\end{equation}
\begin{equation}
\label{e6} \ddot{\chi}  + 3H \dot{\chi} +  2 \frac{\dot{P}}{P}
\dot{\chi} - \frac{P_{,\chi}}{P} \dot{\chi} ^2 -
\frac{TT_{,\chi}}{P^2} \dot{\varphi} ^2 + \frac{U_{,\chi}}{P^2} =
0,
\end{equation}

\noindent where $H \equiv \frac{\dot{a}}{a}$ is the Hubble rate.

\noindent Following the papers \cite{Nibbelink,GWBM}, we define
the new "adiabatic" field $\sigma$ such that
\begin{equation}
\label{e7} \dot{ \sigma }= T\dot{ \varphi } \cos \Theta  + P \dot{
\chi} \sin \Theta ,
\end{equation}

\noindent where
\begin{equation} \label{e8}
\frac{T\dot{\varphi}}{\sqrt{T^2\dot{\varphi} ^2 \!+\! P^2
\dot{\chi} ^2}} \!=\! \cos \Theta ,~
\frac{P\dot{\chi}}{\sqrt{T^2\dot{\varphi} ^2 \!+\! P^2 \dot{\chi}
^2}} \!=\! \sin \Theta.
\end{equation}

\noindent Using definitions
\begin{eqnarray}
\label{e9}  U_{\sigma} = \frac{U_{,\varphi}}{T} \cos \Theta  +
\frac{U_{,\chi}}{P} \sin \Theta, \nonumber \\
U_{s} = \frac{U_{,\chi}}{P}\cos \Theta - \frac{U_{,\varphi}}{T}
\sin \Theta  ,
\end{eqnarray}

\noindent from the chiral field equations (\ref{e5}) and
(\ref{e6}) we can write very useful relations:
\begin{equation}
\label{e10}  \ddot{\sigma}  + 3H \dot{\sigma} + U_{\sigma} = 0,
\end{equation}
\begin{equation}
\label{e11}  \dot{\Theta} = - \frac{U_s}{\dot{\sigma}} -
\frac{P_{,\varphi}}{T} \dot{\chi} +\frac{T_{,\chi}}{P}
\dot{\varphi}.
\end{equation}

\section{Perturbations.}

We investigate scalar linear perturbations about the spatially
flat FRW background. The  corresponding perturbed line element can
in general be written  as \cite{MFB, LL}
\begin{eqnarray}
\label{e12}ds^2 =  (1+2\phi)dt^2 - 2aB_{,i}dx^idt ~~~~~~~~~~\nonumber \\
-a^2\left[ (1-2\psi)\delta_{ij} + 2E_{,ij}\right] dx^idx^j.
\end{eqnarray}

The non-linear sigma model components can also be decomposed into
a homogeneous background fields $ \varphi (t) $, $ \chi (t) $ and
small inhomogeneities $ \delta\varphi $, $\delta\chi$:
\begin{eqnarray}
\varphi({\mathbf{x}}, t) = \varphi(t) + \delta \varphi
({\mathbf{x}}, t), \nonumber \\
\chi({\mathbf{x}}, t) = \chi(t) + \delta \chi({\mathbf{x}}, t).
\end{eqnarray}

Gauge transformations $\tilde{t} =t + \xi ^0 (t, {\mathbf{x}})$,
$\tilde{{\mathbf{x}}} ={\mathbf{x}} + \xi ^ {,i} (t, {\mathbf{x}})
$, where $ \xi ^0 (t, {\mathbf{x}}) $, $ \xi (t, {\mathbf{x}}) $
are the arbitrary scalar functions, allow one to simplify the
metric tensor by a suitable choice of constant-time hypersurfaces
and spatial coordinates inside them. Here we operate without
specified any particular gauge conditions.

Using perturbations of the energy-momentum tensor $ \delta T ^
{\mu} _ {\nu} $ and those of the Einstein tensor $ \delta G ^
{\mu} _ {\nu} $\cite {MFB}, one can obtain the perturbed Einstein
equations for inhomogeneities with the comoving wave number
$\mathbf{k}$:
\begin{eqnarray}
\label{e13} 3H(H\phi_{\mathbf{k}} + \dot{\psi_{\mathbf{k}}}) +
\frac{k^2}{a^2} \left[ \psi_{\mathbf{k}}+ H(a^2 \dot{E}
_{\mathbf{k}} - aB_{\mathbf{k}}) \right] \nonumber\\
= -\frac{\kappa^2}{2} \delta \rho _{\mathbf{k}},
\end{eqnarray}
\begin{equation}
\label{e14} \dot{\psi}_{\mathbf{k}}+ H \phi_{\mathbf{k}} = -
\frac{\kappa^2}{2} \delta q_{\mathbf{k}},
\end{equation}
\begin{eqnarray}
\label{e15} \ddot{\psi}_{\mathbf{k}} + 3H \dot{\psi}_{\mathbf{k}}
+ H \dot{\phi}_{\mathbf{k}} + (2\dot{H} + 3H^2) \phi_{\mathbf{k}}
\nonumber \\
= \frac{\kappa^2}{2}({\delta\rho_{\mathbf{k}} -2U _ {,\varphi}
\delta \varphi_{\mathbf{k}} - 2U_{,\chi} \delta\chi_{\mathbf{k}}
}),
\end{eqnarray}
\begin{equation}
\label{e16} \phi_{\mathbf{k}}\! - \psi_{\mathbf{k}}\! +
H(aB_{\mathbf{k}}\! - a^2\dot{E}_{\mathbf{k}}) +
(aB_{\mathbf{k}}\! - a^2\dot{E}
_{\mathbf{k}})^{{\displaystyle{\cdot}}}\!=0,
\end{equation}

\noindent where the quantities $ \delta\rho_{\mathbf{k}} $, $
\delta q_{\mathbf{k}} $ are
\begin{eqnarray}
\label{e17} \delta\rho_{\mathbf{k}} = T^2 \dot{\varphi} \delta
\dot{\varphi} _{\mathbf{k}} +(U_{,\varphi} + PP_{,\varphi}
\dot{\chi}^2 + TT_{,\varphi} \dot{\varphi}^2)
\delta\varphi _{\mathbf{k}} \nonumber \\
+ P^2 \dot{\chi} \delta\dot{\chi}_{\mathbf{k}}\!+\! (U_{,\chi}
\!+\! PP_{,\chi}\dot{\chi}^2 \!+\! TT_{,\chi} \dot{\varphi}^2)
\delta\chi_{\mathbf{k}} \!-\! \dot{\sigma}^2 \phi_{\mathbf{k}} ,
\end{eqnarray}
\begin{equation}
\label{e18} \delta q_{\mathbf{k}} = - T^2\dot{\varphi} \delta
\varphi _{\mathbf{k}}  - P^2 \dot{\chi} \delta\chi_{\mathbf{k}} .
\end{equation}

In studying the evolution of perturbations, gauge-invariant
quantities are very convenient. The following gauge-invariant
quantities are often used:
\begin{eqnarray}
\Phi = \phi + \left(aB - a^2\dot{E} \right)^
{\displaystyle{\cdot}}, \nonumber \\
\label{e19} \Psi = \psi - H\left( aB - a^2\dot{E} \right).
\end{eqnarray}

\noindent They were constructed by Bardeen \cite{Bardeen} and are
widely used in the influential report \cite{MFB}. The comoving
curvature perturbation ${\cal R}_ {\mathbf{k}}$ can be expressed
using the quantities $\Phi_{\mathbf{k}} $ and $\Psi_{\mathbf{k}}$
as follows \cite {Lyth85}:
\begin{equation}
\label{e20} {\cal R}_{\mathbf{k}}=\Psi_{\mathbf{k}} -
\frac{H(H\Phi_{\mathbf{k}} +\dot{\Psi}_{\mathbf{k}})} {\dot{H}}.
\end{equation}

Eq. (\ref{e16}) can be rewritten in the simple form
\begin{equation}
\label{e21} \Phi_{\mathbf{k}}=\Psi_{\mathbf{k}}.
\end{equation}

Taking a time derivative of Eq. (\ref{e20}), using the background
field equations and Eqs. (\ref{e13}) - (\ref{e16}), we find the
equation of motion of the comoving curvature perturbation
\begin{eqnarray}
\label{e22} \dot{\cal R}_{\mathbf{k}} = 2H\dot{\varphi}\dot{\chi}
\frac{T^2\dot{\varphi}U_{,\chi}-P^2\dot{\chi} U_{,\varphi}}
{(T^2\dot{\varphi}^2+ P^2\dot{\chi}^2)^2} \left(\frac{ \delta
\varphi_{\mathbf{k}}}{\dot{\varphi}} -\frac{\delta \chi_
{\mathbf{k}}}{\dot{\chi}}\right)\nonumber \\
+ \frac{H}{\dot{H}} \frac{k^2}{a^2} \Psi_{\mathbf{k}}.
\end{eqnarray}

In the description of the evolution of the inhomogeneities,  we
shall also use the new variables $\delta s $ and $ \delta\sigma $
\cite {Nibbelink}:
\begin{eqnarray}
\label{e24} \delta s =P \cos \Theta \delta \chi - T\sin \Theta
\delta \varphi ,\nonumber \\
\delta \sigma =P \sin \Theta \delta \chi + T\cos \Theta \delta
\varphi.
\end{eqnarray}

The quantities $\delta\sigma$ and $\delta s$ were termed in Ref.
\cite{GWBM}  "adiabatic" and "entropy" perturbations. As
indicated, by the name, perturbations with $ \delta s =0 $ are
only adiabatic. By definition, it is clear that $ \delta \sigma $
is associated with inhomogeneities of the field $\sigma$ while $
\delta s $ corresponds to inhomogeneities of the entropy field $s$
defined by
\begin{equation}
\label{e25}d s = P \cos \Theta d \chi - T\sin \Theta d \varphi.
\end{equation}
In some two-field inflationary models \cite{PSt94_}, there is a
very simple relationship between the generalized entropy
perturbations $\delta s$ after inflation and the specific entropy
at the start of radiation dominated stage \cite{GWBM,Langlois}.

Using variables $ \delta\sigma $ and $\delta s $, Eq. (\ref{e22})
can be reduce to
\begin{equation}
\label{e26} \dot{\cal R}_{\mathbf{k}} = \frac{H}{\dot{H}}
\frac{k^2}{a^2} \Psi_{\mathbf{k}} -2H\frac{U_s}{\dot{\sigma}
^2}\delta s_{\mathbf{k}},
\end{equation}

\noindent which formally coincides with the corresponding
equations of Refs. \cite{GWBM}, \cite{MFB02}. For purely adiabatic
perturbations ($ \delta\varphi / \dot {\varphi} = \delta\chi /
\dot {\chi} $), we have $\delta s =0$, and at large scales ($ k /
a \ll H$), a conservation law is valid for  comoving curvature
perturbation ${\cal R}_{\mathbf{k}}$.

The perturbed chiral field equations also can be presented in the
terms of the new variables. Directly perturbing the chiral field
equations \cite {Chervon02} for sigma model (\ref{e1}), we find:
\begin{eqnarray}
\label{e27} \ddot{\delta\varphi}_{\mathbf{k}} \!+\! \left( \!3H
\!+ 2 \frac{\dot{T}}{T}\! \right)\! \dot{\delta\varphi}_
{\mathbf{k}} \!+ 2 \!\left(\! \frac{ T_{,\chi}}{T}\dot{\varphi}
\!- \frac{PP_{,\varphi}}{T^2} \dot{\chi} \!\right)\!
\dot{\delta\chi}_{\mathbf{k}}                    \nonumber \\
+\! \left(\! 2 \frac{(a^3TT_{,\varphi} \dot{\varphi}\dot{)}}{a^3
T^2} \!-\! \frac{(T^2)_{,\varphi\varphi} }{2T^2} \dot{\varphi} ^2
\!-\! \frac{(P^2)_{,\varphi \varphi}}{2T^2} \dot{\chi} ^2
\!\right)\!\delta\varphi_{\mathbf{k}}           \nonumber \\
+ \!\left(\!2\frac{(a^3TT_{,\chi} \dot{\varphi}\dot{)}}{a^3 T^2}
\!-\! \frac{(T^2)_{,\varphi \chi}}{2T^2} \dot{\varphi} ^2 \!-\!
\frac{(P^2)_{,\varphi \chi} }{2T^2} \dot{\chi} ^2\!\right)\!
\delta\chi_{\mathbf{k}}                         \nonumber \\
+ \left(\frac{k^2}{a^2}+ \frac{U_{,\varphi \varphi}}{T^2}\right)
\delta \varphi _{\mathbf{k}} +\frac{U_{,\varphi \chi}}{T^2}
\delta\chi_{\mathbf{k}}+ 2\frac {U_{,\varphi}} {T^2}
\phi_{\mathbf{k}}                               \nonumber \\
=  \dot{\varphi} \left( \dot{\phi}_{\mathbf{k}} + 3
\dot{\psi}_{\mathbf{k}} -\frac{k^2B_{\mathbf{k}}}{a}
+k^2\dot{E}_{\mathbf{k}}\right)\!,
\end{eqnarray}
\begin{eqnarray}
\label{e28}  \ddot{\delta\chi}_{\mathbf{k}}  \!+\! \left(\! 3H
\!+\! 2 \frac{\dot{P}}{P} \!\right)\! \dot{\delta\chi}_
{\mathbf{k}} \!+ 2\!\left(\! \frac{P_{,\varphi}}{P} \dot{\chi} \!-
\frac{TT_{,\chi}} {P^2}\dot{\varphi}\!\right) \!
\dot{\delta\varphi}_{\mathbf{k}}  \nonumber \\
+\! \left(\!2 \frac{(a^3PP_{,\chi} \dot{\chi}\dot{)}}{a^3 P^2}
\!-\! \frac{(P^2)_{,\chi \chi}}{2P^2} \dot{\chi} ^2 \!-\!
\frac{(T^2)_{,\chi\chi}}{2P^2} \dot{\varphi} ^2 \!\right)\!
\delta\chi_{\mathbf{k}}               \nonumber \\
+\! \left({ \!2 \frac{(a^3PP_{,\varphi} \dot{\chi}\dot{)}}{a^3
P^2} \!-\! \frac{(P^2)_{,\varphi \chi}}{2P^2} \dot{\chi} ^2 \!-\!
\frac{(T^2)_{,\varphi \chi}}{2P^2}\dot{\varphi} ^2 }\!\right)\!
\delta \varphi_{\mathbf{k}}       \nonumber \\
+ \left(\frac{k^2}{a^2}+ \frac{U_{,\chi \chi}}{P^2}\right) \delta
\chi_{\mathbf{k}} + \frac{U_{,\varphi \chi}}{P^2} \delta \varphi_
{\mathbf{k}} + 2\frac{U_{,\chi}}{P^2} \phi_{\mathbf{k}}      \nonumber \\
= \dot{\chi} \left( \dot{\phi}_{\mathbf{k}} + 3
\dot{\psi}_{\mathbf{k}} - \frac{k^2 B_{\mathbf{k}}}{a} + k ^2
\dot{E}_{\mathbf{k}}\right) .
\end{eqnarray}

After some calculations, the following equations for the
inhomogeneities $\delta\sigma _{\mathbf{k}}$ and  $\delta s _
{\mathbf{k}}$ can be derived:
\begin{eqnarray}
\label{e29}\ddot{\delta\sigma}_{\mathbf{k}} \!+\! 3H\dot{\delta
\sigma} _{\mathbf{k}} \!+\! \left(\!\frac{k^2}{a^2} \!+\!
\frac{U_{s}} {\dot{\sigma}} \dot{\Theta} \!+\! U_{\sigma\sigma}
\!-\! \frac{U_{,\chi}} {\dot{\sigma}}\frac{\dot{P}}{P^2}
\sin\Theta \right.\nonumber \\
 \left. - \frac{U_{,\varphi}} {\dot{\sigma}}\frac{\dot{T}}{T^2}
\cos\Theta \!\right)\! \delta\sigma_{\mathbf{k}} +
2\left(\!\frac{U_{s}} {\dot{\sigma}}\delta s _{\mathbf{k}}\!
\right) ^{\displaystyle {\cdot}} \!- 2 \frac{U_{\sigma}}
{\dot{\sigma}}
\frac{U_s} {\dot{\sigma}} \delta s _{\mathbf{k}}   \nonumber \\
=-2U_{\sigma} \phi _{\mathbf{k}} + \dot{\sigma}\left(
\dot{\phi}_{\mathbf{k}}+ 3 \dot{\psi} _{\mathbf{k}} + k ^2
\dot{E}_{\mathbf{k}} - \frac{k^2B_{\mathbf{k}}}{a} \right)\!,
\end{eqnarray}
\begin{eqnarray}
\label{e30}\ddot{\delta s}_{\mathbf{k}}+ 3H\dot{\delta s}_
{\mathbf{k}} +\left(\frac{U_{\sigma}} {PT}\left( P_{,\varphi}\cos
\Theta+ T_{,\chi}\sin \Theta\right)  \right.\nonumber \\
+\left( \frac{P_{,\varphi} T_{,\varphi}}{T^2} + \frac{P_{,\chi}
T_{,\chi}} {P^2} - \frac{P_{,\varphi\varphi}}{T} - \frac
{T_{,\chi\chi}}{P}\right)\frac{\dot{\sigma} ^2}{PT}\nonumber \\
+ \frac{U_{,\chi}}{P^2}\cos \Theta \left( \frac{P_{,\varphi}}{T}
\sin \Theta - \frac{P_{,\chi}}{P} \cos \Theta\right) \nonumber\\
- \frac{U_{,\varphi}}{T^2}\sin \Theta \left( \frac{T_{,\varphi}}
{T}\sin \Theta - \frac{T_{,\chi}}{P} \cos \Theta\right)
 \nonumber \\
\left.+ \frac{k^2 }{a^2} +3\left(\frac{U_{s}} {\dot{\sigma}}
\right) ^{2}+ U_{ss}  \right)\delta s_{\mathbf{k}} = 2\frac{U_{s}}
{\dot{\sigma} ^2} \epsilon _m,
\end{eqnarray}

\noindent where
\begin{equation}
\label{e31}U_{ss}= \frac{U_{,\chi\chi}}{P^2}\cos ^2 \Theta -
\frac{U_{,\varphi\chi}}{PT}\sin 2\Theta  +
\frac{U_{,\varphi\varphi}}{T^2}\sin ^2 \Theta,
\end{equation}
\begin{equation}
\label{e32}U_{\sigma\sigma}= \frac{U_{,\chi\chi}}{P^2}\sin ^2
\Theta +  \frac{U_{,\varphi\chi}}{PT} \sin 2\Theta  +
\frac{U_{,\varphi\varphi}}{T^2} \cos ^2 \Theta.
\end{equation}

\noindent Here we use the gauge-invariant quantity $ \epsilon _m
\equiv \delta\rho_{\mathbf{k}} -3H\delta q_{\mathbf{k}} $
\cite{Bardeen}, the comoving density perturbation:
\begin{equation}
\label{e33} \epsilon _m  = \dot{\sigma} \left[  \dot{ \delta
\sigma _{\mathbf{k}}}+2\frac{U_s}{\dot{\sigma}}\delta
s_{\mathbf{k}} - \dot{\sigma} \phi _{\mathbf{k}}-
\frac{\ddot{\sigma}} {\dot{\sigma}} \delta\sigma_{\mathbf{k}}
\right].
\end{equation}

\noindent Since from the Eqs. (\ref{e13}) and (\ref{e14}) we have
\begin{equation}
\label{e34}\frac{k^2}{a^2}\Psi _{\mathbf{k}} = -\frac{\kappa
^2}{2}\epsilon _m,
\end{equation}
at large scales ($k / a \ll H$) the right hand side of Eq.
(\ref{e30}) becomes negligible. As follows from Eq. (\ref{e30}),
the large-scale entropy perturbations are decoupled from adiabatic
and metric perturbations, which generalizes the conclusion of Ref.
\cite{GWBM} on inflationary models with the action (\ref{e1}).

Eq. (\ref{e29}) can be rewritten as an equation for
Sasaki-Mukhanov gauge-invariant variable
\begin{equation}
\label{e35}Q=\delta\sigma +\frac{\dot{\sigma}}{H}\psi.
\end{equation}

\noindent Using this quantity, one can obtain:
\begin{eqnarray}
\label{e36}\ddot{Q}_{\mathbf{k}}+ 3H\dot{Q}_{\mathbf{k}} +
\left(\frac{k^2}{a^2} + \frac{U_{s}} {\dot{\sigma}}\dot{\Theta} -
\frac{U_{,\chi}}{\dot{\sigma}} \frac{\dot{P}}{P^2} \sin\Theta
\right.\nonumber \\
\left.- \frac{U_{,\varphi}}{\dot{\sigma}}\frac{\dot{T}}{T^2}
\cos\Theta - \frac{\kappa
^2}{a^3}\left(\frac{a^3\dot{\sigma}^2}{H}\right)^
{\displaystyle{\cdot}}+ U_{\sigma\sigma} \!\right)\!
Q_{\mathbf{k}}  \nonumber \\
=- 2\left(\frac{U_{s}} {\dot{\sigma}}\delta s_{\mathbf{k}} \right)
^ {\displaystyle{\cdot}} + 2 \left(\frac{U_{\sigma}}{\dot{\sigma}}
+ \frac{\dot{H}}{H}\right)\frac{U_s} {\dot{\sigma}} \delta
s_{\mathbf{k}} .
\end{eqnarray}

Eqs. (\ref{e30}) and (\ref{e36}) in specific case $T^2=1$, $P^2 =
P^2(\varphi)$ coincide with the corresponding equations of Ref.
\cite{MFB02}, and in the  slow roll conditions reduce to equations
of Ref. \cite{Nibbelink}.

\section{Application to slow roll inflationary models.}

Let us consider an application of the above expressions to slow
roll inflationary models. Quantum vacuum fluctuations of chiral
fields become classical inhomogeneities after they leave the
horizon. The Fourier component of chiral fields perturbations at
horizon crossing  are given by \cite{GW96}
\begin{eqnarray}
\label{e37} \delta \varphi _{\mathbf{k}} (t_{\mathbf{k}})= \frac
{1}{T(t_{\mathbf{k}})} \frac {{H (t_{\mathbf{k}})}} {\sqrt {2k^3}}
e _ {\varphi} ({\mathbf{k}}) ,\\
\delta \chi _{\mathbf{k}} (t_{\mathbf{k}}) = \frac
{1}{P(t_{\mathbf{k}})} \frac {H (t_{\mathbf{k}})} {\sqrt {2k^3}} e
_ {\chi} ({\mathbf{k}}).
\end{eqnarray}

\noindent Here $t_{\mathbf{k}}$ is the instant of horizon crossing
($a(t_{\mathbf{k}}) H(t_{\mathbf{k}}) $ $= k$), and $e_{\varphi}$,
 $e _{\chi}$ are real Gaussian stochastic variables with the
following properties:
\begin{eqnarray}
\label{e38} \left\langle {e _ {\varphi} (\mathbf{k}) e _ {\chi}
(\mathbf{k} ')} \right\rangle = \delta _ {\varphi\chi} \delta ^
{(3)} \left (\mathbf{k} - \mathbf{k} ' \right),\\
\left\langle {e _ {\phi} (\mathbf{k})} \right\rangle =
\left\langle {e _ {\chi} (\mathbf{k})} \right\rangle =0.
\end{eqnarray}

From (\ref{e24}) and (\ref{e37}) one can derive the following
expressions for $\delta\sigma_{\mathbf{k}}$ and $\delta
s_{\mathbf{k}}$:
\begin{equation}
\label{e39}\delta\sigma _{\mathbf{k}} (t_{\mathbf{k}})= \frac {{H
(t_{\mathbf{k}})}} {\sqrt {2k^3}} e _{\sigma} ({\mathbf{k}}) ,~
\delta s _{\mathbf{k}}(t_{\mathbf{k}}) = \frac {H
(t_{\mathbf{k}})} {\sqrt {2k^3}} e _{s} (\mathbf{k}),
\end{equation}

\noindent where $e _{\sigma}$, $e _{s}$ are independent Gaussian
random quantities with zero average and unit dispersion.

Being restricted to the most important non-decreasing modes of
long-wavelength inhomogeneities, as well as in case of
single-field inflationary models, it is possible to write the
inequality:
\begin{equation}
\label{e40}\dot{\Phi}_{\mathbf{k}} \ll H\Phi_{\mathbf{k}}.
\end{equation}

Another standard assumption is that, for non-decreasing modes in
the slow-roll regime $|\ddot{\delta\varphi}/\dot{\delta\varphi}|
\ll 3H$, $|\ddot{\delta\chi}/\dot{\delta\chi}| \ll 3H$ and
therefore \cite{GWBM}, \cite{Nibbelink}
\begin{equation}
|\ddot{\delta \sigma}| \ll 3H |\dot{\delta
\sigma}|,~~~~~~~~~~|\ddot{\delta s}| \ll 3H |\dot{\delta s}|.
\end{equation}

This allows us to disregard  $\ddot{\delta \sigma}$ and
$\ddot{\delta \sigma}$ in the perturbed field equations
(\ref{e29}), (\ref{e30}). Thus, Eq. (\ref{e30}) may be written as
\begin{equation}
\label{en_slow}\dot{\delta s}_{\mathbf{k}} = A(t) \delta
s_{\mathbf{k}},
\end{equation}

\noindent where
\begin{eqnarray}
A(t)=-\frac{1}{3H}\left(\frac{U_{\sigma}} {PT}\left( P_{,\varphi}
\cos\Theta+ T_{,\chi}\sin \Theta\right)\right.  \nonumber \\
+ 3\left(\frac{U_{s}} {\dot{\sigma}} \right)^{2}+ \frac{U_{,\chi}}
{P^2}\cos \Theta \left( \frac{P_{,\varphi}}{T}\sin \Theta
- \frac{P_{,\chi}}{P} \cos \Theta\right)\nonumber\\
+ U_{ss} - \frac{U_{,\varphi}}{T^2}\sin \Theta \left( \frac
{T_{,\varphi}} {T}\sin \Theta - \frac{T_{,\chi}}{P} \cos \Theta
\right)  \nonumber \\
\left.+\left( \frac{P_{,\varphi} T_{,\varphi}}{T^2} + \frac
{P_{,\chi} T_{,\chi}} {P^2} - \frac{P_{,\varphi\varphi}}{T} -
\frac {T_{,\chi\chi}}{P}\right)\frac{\dot{\sigma} ^2}{PT}\right).
\end{eqnarray}

\noindent Taking into account the inequality (\ref{e40}),
equations (\ref{e14}) and (\ref{e20}) can be simplified:
\begin{equation}
\label{e42}\Phi _{\mathbf{k}}(t)=\frac{\kappa ^2}{2H} \dot{\sigma}
\delta\sigma _{\mathbf{k}},
\end{equation}
\begin{equation}
\label{e43}{\cal R}_{\mathbf{k}}= -\frac{H^2}{\dot{H}}
\Phi_{\mathbf{k}}.
\end{equation}

Integrating Eq. (\ref{e26}), one can obtain the following
expression for the long-wavelength comoving curvature perturbation
${\cal R}_{\mathbf{k}}$ :
\begin{equation}
{\cal R}_{\mathbf{k}}(t) = {\cal R}_{\mathbf{k}} (t_{\mathbf{k}})
-  \frac {H (t_{\mathbf{k}})} {\sqrt {2k^3}} J(t,t_{\mathbf{k}}) e
_{s} ({\mathbf{k}}),
\end{equation}

\noindent where
\begin{equation}
J(t,t_{\mathbf{k}}) =\!\!\int _{t_{\mathbf{k}}} ^t
\!\frac{2HU_s}{\dot{\sigma} ^2} \frac{\delta s}{\delta s
(t_{\mathbf{k}})} dt=\!\!\int _{t_{\mathbf{k}}} ^t
\!\frac{2HU_s}{\dot{\sigma} ^2} \alpha_{\mathbf{k}}(t) dt.
\end{equation}

Here the function $ \alpha _ {\mathbf {k}} (t) $ is a solution of
Eq. (\ref{en_slow}), normalized by the condition $ \alpha _
{\mathbf {k}} ( t _ {\mathbf {k}}) =1 $. Using first of relations
(\ref{e39}), background equation (\ref{e4}) and also equations
(\ref{e42}), (\ref{e43}) one can therefore write:
\begin{equation}
\label{e44}{\cal R}_{\mathbf{k}}(t) = \frac {H(t_{\mathbf{k}})}
{\sqrt {2k^3}}\!\left(\!\frac{ H(t_{\mathbf{k}})} {\dot{\sigma}
(t_{\mathbf{k}})} e _ {\sigma} ({\mathbf{k}}) -
J(t,t_{\mathbf{k}})e_{s} ({\mathbf{k}})\!\right)\! .
\end{equation}

The power spectrum \cite{LL} can now be written  in the following
form:
\begin{equation}
\label{e45}{\cal P}_{\cal R}(t)= \frac{H ^2(t_{\mathbf{k}})}
{4\pi^2} \left(\frac{H^2(t_{\mathbf{k}})}{\dot{\sigma}^2
(t_{\mathbf{k}})}+  J^2(t,t_{\mathbf{k}})\right) .
\end{equation}

This formula is valid also after the slow-roll stage if,  for
finding the function $ \alpha _ {\mathbf {k}} (t) $   a
long-wavelength limit of Eq. (\ref{e30}) is used instead of Eq.
(\ref{en_slow}). If, from certain time, the entropy perturbation
$\delta s$ is so small, that the comoving curvature perturbation
${\cal R}$ remains actually constant, the expression obtained
above describe the adiabatic power spectrum.

A commonly used characteristic of a power spectrum is its spectral
index $n_{s}$ \cite{LL}. Using (\ref{e45}) and background
equations one can conclude, that the spectral index for
inflationary models under consideration is given by:
\begin{eqnarray}
\label{e46}n_s -1\equiv \frac{\partial \ln {\cal P}_{\cal
R}(t)}{\partial \ln k} = \frac{1}{H}\frac{\partial \ln {\cal
P}_{\cal R}(t)}{\partial t_{\mathbf{k}}}
=2\frac{\dot{H}(t_{\mathbf{k}})}
{H^2(t_{\mathbf{k}})}\nonumber \\
- \frac{\kappa ^2 + \frac{2H(t_{\mathbf{k}})\ddot{\sigma}
(t_{\mathbf{k}})} {\dot{\sigma}^3 (t_ {\mathbf{k}})}  + 2
\left(\frac{2U_s (t_{\mathbf{k}})} {\dot {\sigma} ^2
(t_{\mathbf{k}})}J + \frac{A(t_{\mathbf{k}})}{H(t_{\mathbf{k}})}
J^2\right) } {\frac{H^2 (t_{\mathbf{k}} )} {\dot{\sigma}^2
(t_{\mathbf{k}})} +  J^2}.
\end{eqnarray}

The spectral index is time-dependent due the time-dependence of
the quantity $J(t,t_{\mathbf{k}})$.

Eqs. (\ref{e45}), (\ref{e26}), (\ref{e30}) and (\ref{e36}) are the
main results of this work. These equations are very well suitable
for numerical calculations of  perturbation evolution in
inflationary models based on non-linear sigma models.

\subsection{Numerical example.}
As an example, we shall describe an inflationary model with a
single massive scalar field and additional higher order curvature
terms due to curvature squared. The corresponding action looks
like:
\begin{eqnarray}
\label{e47}S = \int d^4 x\sqrt { |\tilde{g}|} \left\{- \frac
{\tilde{R}} {2 \kappa^2}+ \frac{\tilde{R}^2}{2 \kappa^2 M^2}
\right. \nonumber \\
\left.+ \frac{1}{2} \varphi_{;\mu} \varphi_{;\nu} \tilde{g}^
{\mu\nu} - \frac{1}{2} m^2 \varphi ^2 \right\}.
\end{eqnarray}

\begin{figure}
\centering
\includegraphics[width=0.47\textwidth]{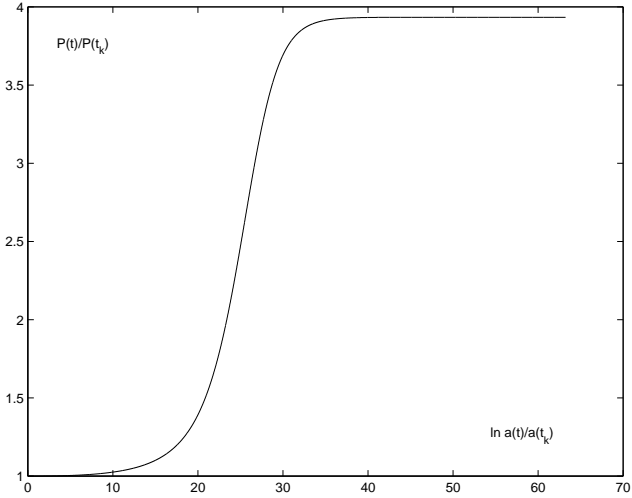}
{\caption{The time dependence of ${\cal P}(t)/{\cal
P}(t_{\mathbf{k}})$ with $M/ m =5$,
$\varphi(t_{\mathbf{k}})=14/\kappa$,
$\chi(t_{\mathbf{k}})=4/\kappa$.}}
\end{figure}

\noindent Making the conformal transformation $g_{ij} = \Omega ^2
\tilde {g} _ {ij} $ with $ \Omega^2 = 1 -\frac{\tilde {R} ^2}
{3M^2}$, one gets the action in the Einstein frame:
\begin{eqnarray}
\label{e48}S =\!\! \int \!d^4 x\sqrt { |g|} \!\left\{ \!
\frac{-R}{2 \kappa^2} \!+\! \frac{1}{2} \chi_{;\mu}\chi^{;\mu}
\!\!+\! \frac{e^{-\sqrt{\frac{2}{3}} \kappa \chi}}{2}
 \varphi_{;\mu}\varphi^{;\mu}\right. \nonumber \\
\left. - \frac{3}{4}\frac{M^2}{\kappa^2} \left[ 1 \!-\!
e^{-\sqrt{\frac{2}{3}} \kappa \chi}\right] ^2 \!-\! \frac{1}{2}
e^{-2 \sqrt{\frac{2}{3}} \kappa \chi} m^2 \varphi^2 \!\right\} \!,
\end{eqnarray}

\noindent where the new field $\chi\equiv \sqrt {3/ 2}
{\kappa}^{-1} \ln \left (1-{\tilde{R} ^2}/{3M^2}\right) $ is
introduced. Note that when $\chi \ll \kappa ^ {-1}$, then
$U(\varphi,\chi) \approx M^2\chi^2 /2+ m^2\varphi^2/2$. Some
inflationary models more general than (\ref{e47})  were considered
earlier in the preheating context in \cite{TsMT99, StTY01} (These
papers  took into account not only higher curvature terms, but
also a possible  non-minimal coupling $\xi \varphi ^2 \tilde {R}
$).

We are interesting here in the case when the chiral fields $
\varphi $ and $ \chi $ are comparable in magnitude and are both in
a slow-roll regime at the beginning of the inflationary stage. In
the analysis of the background and inhomogeneities evolution, we
shall be restricted to the parameter area $M > m $. Since the
field $ \chi $ is heavier, it breaks up faster than field
$\varphi$, and soon after the inflation influence of the field $
\chi $ on the scale factor dynamic and evolution of
inhomogeneities is vanishingly small. Besides, after inflation, $
\Omega ^2 \cong 1 $, therefore actions in the original frame
(\ref{e47}) and in the Einstein frame (\ref{e48}) almost do not
differ. Accordingly, the inhomogeneities calculated in both frames
after inflation asymptotically coincide. It is clear that the
inhomogeneities in this model after the inflationary stage are to
a great extent adiabatic.

\begin{figure}
\centering
\includegraphics[width=0.47\textwidth]{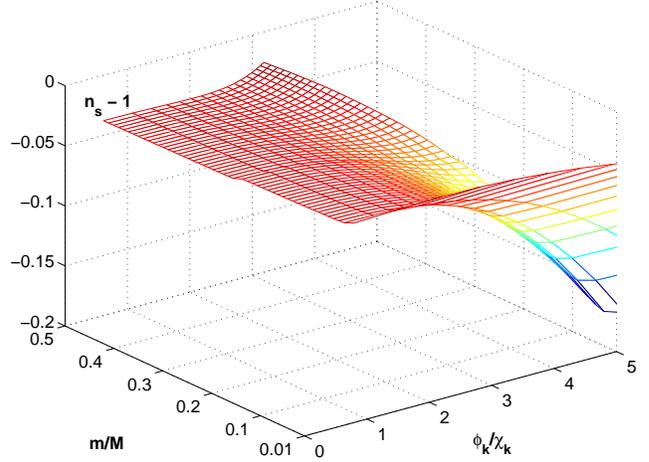}
{\caption{Numerical simulations of the spectral index after
inflation. Here it is assumed that the perturbations leave the
horizon 60 e-folds before the end of inflation; the inflation ends
when $\dot{H}/3H^2 =1 $.}}
\end{figure}

The background and perturbed equations for the sigma model
(\ref{e48}) can be solved by numerical simulations. For  finding
the long-wavelength comoving curvature perturbation ${\cal R}$
generated on inflationary stage, it is very convenient to use Eqs.
(\ref{e26}) and (\ref{e30}). Eq. (\ref{e30}) is much easier to
solve than the initial set of perturbed Einstein and field
equations. It can be made by using standard packages of numerical
computations. The results of the calculations are presented in
Figs. 1 and 2. These pictures show the time dependence ${\cal P}
(t) / {\cal P}(t_{\mathbf{k}})$ for a specific choice of
parameters and the dependence of the spectral index on the model
parameters.

\subsection{Acknowledgement.} {I thank S.V. Chervon and V.M. Zhuravlev for attention to
the work and useful discussions.}

\end {document}